*Original Article*

# A Secure Third-Party Auditing Scheme Based on Blockchain Technology in Cloud Storage

S. M. Udhaya Sankar[1], D. Selvaraj[2], G.K. Monica[3], Jeevaa Katiravan[4]

[1,3]Department of Information Technology, Velammal Institute of Technology, Chennai, India
[2]Department of Electronics and Communication Engineering, Panimalar Engineering College, Chennai, India
[4]Department of Information Technology, Velammal Engineering College, Chennai, India

[1]Corresponding Author: udhaya3@gmail.com

*Abstract* - With the help of a shared pool of reconfigurable computing resources, clients of the cloud-based model can keep sensitive data remotely and access the apps and services it offers on-demand without having to worry about maintaining and storing it locally. To protect the privacy of the public auditing system that supports the cloud data exchange system. The data's owner has the ability to change it using the private key and publishes it in the cloud. The RSA Technique is used to produce key codes for the cloud services atmosphere's privacy utilizing the system's baseboard number, disc number, and client passcode for validation. The method is based on a cutting-edge User End Generated (UEG) privacy technique that minimizes the involvement of a third party and improves security checks by automatically documenting destructive activities. To strengthen extensibility, various authorization-assigning modalities and block access patterns were established together with current operational design approaches. In order to meet the demands for decentralization, fine-grained auditability, extensibility, flexibility, and privacy protection for multilevel data access in networked environments, the suggested approach makes use of blockchain technology. According to a thorough performance and security assessment, the current proposal is exceptionally safe and effective.

## 1. Introduction

Data transmission and sharing have expanded recently due to the rapid growth of communications and networking. The need for media content like video, photos, and music is rising. This significant development has made delivering digital technologies services to people and organizations quite expensive [1]. Due to its financial benefits, cloud computing is, in this aspect, an effective and appropriate platform for offering critical IT services. The cloud computing approach serves as the next step in the evolution of a group's IT since it offers a wide range of unrivaled capabilities, including self-service on demand, data services from wherever, quick asset flexibility, geographical autonomy, usage-based billing, and risk assessment.

The positive effects of cloud computing will significantly alter how businesses employ IT. The data in this paradigm is targeted for cloud computing, one of its primary features. From the perspective of users, including information technology firms, adaptable on-request virtual storage systems in the public cloud have many benefits, including reducing the workload for data storage, general data services from various locations, and lower hardware and app costs, preservation, etc. Cloud-based storage will turn data centers into tremendous computational services because of the minimal price and outstanding efficiency of cloud services, which has been the subject of numerous systems' discussions [2-5]. Users will undoubtedly receive excellent web services due to the channel's Trustworthiness and versatile connectivity, as well as the channel's rapid bandwidth expansion.

Almost everyone has data, whether it be public or personal. Because we can only recall some of that knowledge for a short time, we retain backups of collections of information today, frequently in the form of papers, photographs, videos, and music. These days, it is simple to keep such data on standby in the device's backup system, from which it is possible to access it whenever and whenever often is necessary [18-22]. Datastores that can store data to a specific level include hard drives, CD-ROMs, flash drives, and microchips. These gadgets are transportable and may be used wherever to enter any network and acquire the necessary info or data [23-26]. These data can be used, but there is still a chance that doing so will result in devices being lost because of an error delaying access to the data at that point. It was ultimately causing a loss of information and time that prevented us from getting what we needed when we needed it. If the storing device is lost, an unidentified third party could take it and gain access if it is not secured. The third-party might then misuse the data, putting the user in jeopardy.

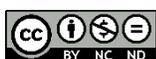

Individuals with technological advances nowadays want things close to hand; they do not wish to waste so much time on anything [27-30]. Thus people try to save time wherever they can. We have all used storage systems, many of which have security features like USB devices, strong passwords, external drives, and much more. The newest tech has also been used to safeguard files [46]. If such safety precautions are adequate for us, do they fully protect our data, fulfill all of our requirements, and if so, how do we ensure it remains safe from cybersecurity threats? Technology has allowed us to post data to a cloud infrastructure service and retrieve it whenever and wherever we need it. Clients can store their information in the cloud, allowing us to retrieve it anywhere globally [32]. In order to give fast information upload and downloading through a safer method, we have some creative ideas here. Keeping a few goals in mind, we focused on the cloud to offer quicker data uploading and downloading times as well as adequate data security [33-35].

The technologies, resources, and apps that seem comparable to those on the web are transformed into the self-service commodity via cloud services. Consumers and programmers are shielded from the operational specifics of cloud services. Applications operate on undetermined physical processes, network management is delegated to third parties, and access permissions are widespread [47]. The realization of computation as effectiveness is a public cloud, which allows users to store private data in the cloud and enables the company's apps and services from a pooling of programmable computer resources on demand. Clients can be spared the responsibility of localized disk space and upkeep by using outsourced info.

Furthermore, because customers no longer have direct access to massive amounts of cloud services, protecting the integrity of data on the internet is a very difficult issue, particularly for individuals with limited computational abilities and resources [37]. Therefore, it is crucial to make possible public audit capability for cloud storage space privacy so that customers can depend on an outside audit agency to bear out the accurateness of cloud services while essential. By combining and pooling ideas, cloud services virtualized systems. Assets are agilely expandable, prices are calculated on a metering basis, and systems and storage may be provided as required from a centralized system [38-40]. It is crucial to make possible public accounting for cloud-based storage space safety so that customers to auditing outside entities to verify the accuracy of cloud services whenever necessary.

## 2. Related Work

The cloud is a comparatively recent communication technology that makes use of several resources. It is a fantastic tool for resource utilization in a business. However, there is a severe security risk. The new incidents on iCloud and Sony Industry's Secret Internal Cloud Breaches demonstrate that privacy remains an essential worry in cloud technology. Privacy in the cloud is not ever guaranteed. So, we looked at the protection mechanisms in place today. We have determined that the current method is typically kept secret, much like when creating OTP. Additionally, it was mentioned that certain cutting-edge businesses use a password vault for their capabilities. Consequently, an extensive algorithm encompassing protection is required, one that takes the shortest amount of time feasible without sacrificing safety.

Guofen Lin et al. [6] offer the key-policy balanced attribute-based cryptography without bilinear pairing calculation for safe cloud-shared data access controls. To explain the various weights assigned to each characteristic, a straightforward weighting process is described. They present a brand-new ABE structure, all without performing bilinear pairing computations. The plan by Lu et al. [7] permitted confidential auditing and included a framework for mobile units to share data via the cloud. Before sharing information with users, their system can verify that authorized people only access it. Additionally, their solution enabled lightweight device activities for both the client and the data requester. Furthermore, unlike the technology, indeed, there did not handle variable data processing (Shen et al. [8]).

Baidaa et al. [9] suggest that an efficient public auditing method for cloud statistics depends on the Boneh-Lynn-Shacham signature to enable public monitoring and uphold data protection. Additionally, batch auditing, as well as data variable, are realized by the suggested system. In order to safeguard the system against unapproved TPA, the suggested system also raises the bar for confirmation with the use of an Automated Blocking Protocol. The challenge of enabling a 3rd party auditor (TPA) to confirm the accuracy of the data saved on the web is one that Dnyanada et al. [10] consider. Using cloud-based storage services that share information with team members is a common practice for users. The present structure thinks that one of the most appealing aspects of cloud services is possibly file sharing between different users. Protecting authenticity and confidentiality from the TPA is a specific issue that arises during the method of public auditors for sharable information stored in the cloud.

Ge et al. [11] provide confidential auditing by advocating for exploring keyword-based investigation using protected dynamic cloud data with symmetric cryptography authentication. To develop an advanced authenticating certificate for a particular keyword, they developed a new method utilizing symmetric-key encryption. In addition, a privacy-preserving dynamic trapdoor hashing authenticity tree was utilized in the Sun et al. [12] technique to encourage community audits and verify data by submitting a trapdoor hash as well as a BLS signature to the Merkle hash tree. They backed up data consistency in their suggested plan but needed more cloud-based data anonymity.

The Cipher - text Attribute-based Encryption method is the emphasis of Nouha Oualha et al. [13]. They encrypt the data while cryptographic keys impose a security and access strategy, ensuring that only authorized data users with the appropriate attributes can decode data. Due to their Processor, storage, power, and other capacity limits, many IoT devices, such as actuators and sensors, cannot be utilized as CP-ABE enforcement endpoints. This study suggests employing efficient pre-computation approaches to expand the fundamental CP-ABE approach. They empirically calculate the potential energy savings afforded by the suggested form of CPABE, demonstrating the use of CP-ABE in the IoT.

To accommodate a range of fine-grained criterion authentication and authorization in multi-authority contexts and lower the computation complexity of key exchange, XiaolongXu et al. [14] introduced the weighting access policy. Furthermore, MPRE-CPABE utilizes proxy re-encryption for connection revocation to reduce computing time. On the Ubuntu and CloudSim platforms, tests are run. According to the outcomes of the experiments, MPRE-CPABE can significantly lower the computing cost of producing critical components and revoke user access rights. Additionally, MPRE-CPABE is demonstrated to be secure using the decision-making bilinear Diffie-Hellman security model (DBDH). Privacy and data must be protected while in the process of auditing in order to keep privacy preservation.

Privacy concerns arise since public validation may be possible. Hence, independent auditors are unable to gain sensitive data about user content. This process is a benefit to retain confidentiality while still being able to confirm the authenticity of data sources because the data holder will not allow the disclosure of his private information to that of an audit firm.

Wang et al. [15], confidentiality technology was created for cloud data storage. When an identifying signal assaults the network, it becomes hazardous, and so this confirmation is demonstrated by the attack's inability to resist. At the same time, numerous methods have been used in cloud technology by Ping et al. [16] to safeguard data privacy. In CP-ABE, Guofeng Lin et al. [17] introduced a cooperative key management mechanism (CKM-CP-ABE). The design achieves decentralized private key creation, issuance, and storage without the need for additional infrastructure. For key updating, a fine-grained and rapid characteristic revocation is offered. The suggested cooperative approach successfully resolves both necessary disclosure and important escrow issues. In the meanwhile, it significantly lowers client decryption latency. According to the assessment, their scheme performs marginally superior to other comparable CP-ABE strategies when it comes to cloud-based outsourcing data exchange on portable devices.

Despite finding several protection philosophies, we conclude there is a slight weakness in the safety precautions. Serious concerns about cloud privacy have been expressed in light of recent iCloud and Sony Internal Storage hacks. In order to deliver a secured Data User Level Key from the client, we have therefore created a new approach by examining the works mentioned above and attempting to achieve a balance between intricacy and consumption of time.

Table 1. Summary of Literature Review

| Year | Author | Techniques | Application | Merits | Demerits |
|---|---|---|---|---|---|
| 2018 | Chen et al. | Blockchain, Smart Contract | Cloud Storage Security | Immutability, Transparency, Accountability | High computation and storage cost |
| 2019 | Zhang et al. | Blockchain, Encryption | Cloud Storage Integrity | Improved data integrity, decentralization | Limited scalability |
| 2020 | Wu et al. | Blockchain, Merkle Tree | Cloud Storage Auditing | Efficiency, Accountability | Limited support for dynamic data |
| 2021 | Huang et al. | Blockchain, Consensus Mechanisms | Cloud Storage Privacy and Security | Strong security guarantees, decentralization | Limited usability, Complex implementation |
| 2022 | Lee et al. | Blockchain, Multi-party Computation | Cloud Storage Compliance | Strong privacy protection, Compliance with regulations | High communication overhead |
| 2017 | Xiong et al. | Blockchain, Proof-of-Retrieve | Cloud Storage Retrieval | Security, Availability, Cost-effectiveness | Limited support for dynamic data |
| 2019 | Cao et al. | Blockchain, Homomorphic Encryption | Cloud Storage Privacy | Strong security guarantees, decentralization | High computational overhead |
| 2020 | Li et al. | Blockchain, Zero-Knowledge Proof | Cloud Storage Access Control | Strong security guarantees, decentralization | High computational overhead |
| 2021 | Zhang et al. | Blockchain, Distributed Storage | Cloud Storage Sharing | Security, Privacy, Decentralization | Limited support for large-scale data sharing |
| 2022 | Wang et al. | Blockchain, Trusted Computing | Cloud Storage Data Protection | Strong security guarantees, Trustworthiness | High complexity of implementation |

Overall, the literature suggests that blockchain-based approaches can enhance various aspects of cloud storage, such as security, privacy, availability, trustworthiness, and decentralization. However, these approaches also have limitations, such as computational overhead, the complexity of implementation, and limited support for certain types of data and applications.

## 3. Objectives

To provide security, use unique identifiers so that we tend to get the user's hard disk number, motherboard number, and some security questions to identify a user uniquely among the others. These attributes set one user apart from others because no two motherboard and hard disc numbers are the same, allowing individuals to stand out from the listing of other customers. The computation for creating a key that uses the hard disc number, motherboard number, client passcodes, and time machine is uncomplicated. However, the logic remains complicated so that attackers could indeed effortlessly attempt to crack the secret message key at a quicker point in time, which is usually offered in two minutes, in order to minimize computational overhead without compromising on security.

The primary goal is to track down harmful auditors by employing automated processes installed at the remote server and watching over all user behavior while logged in. These systems could also characterize whether such a file's contents are faulty or contain many standard errors. An audit is performed on every file using its checksum results; if the file's checksums for uploading and downloading are identical, the client has posted a legal file, may preserve it appropriately, and can retrieve it whenever as well as from wherever. If the file's authentication changes, the client may have submitted a malformed file, or even the file may have become corrupt while being uploaded. Once the client uploads the correct file, the automatic monitoring examines for a matching timestamp and the file's descriptions. If the file qualifies, we may complete the inspection; if the file is damaged, we will see a failed result before we complete the assessment.

## 4. Proposed Methodology

Without downloading the whole set of publicly available data from the cloud, authorized certificates can openly validate the accuracy of the data. Users within the group produce legitimate verification metadata on shared material. During the auditing process, a public validator could tell who signed every piece in the shared resources [41]. The first users are permitted to produce shareable data stored in the cloud, which they then distribute to other users. Group members include both the originating customer and the group users. Each person in the group has access to common data and the capacity to edit the data stream. Both transmitted information and the data needed for its authentication are kept on the remote server. The authenticity of publicly stored data on the remote server is publicly verified by a public validator, such as an independent auditor offering expert data accounting services outside the party expecting to use data access, as shown in Figure 1.

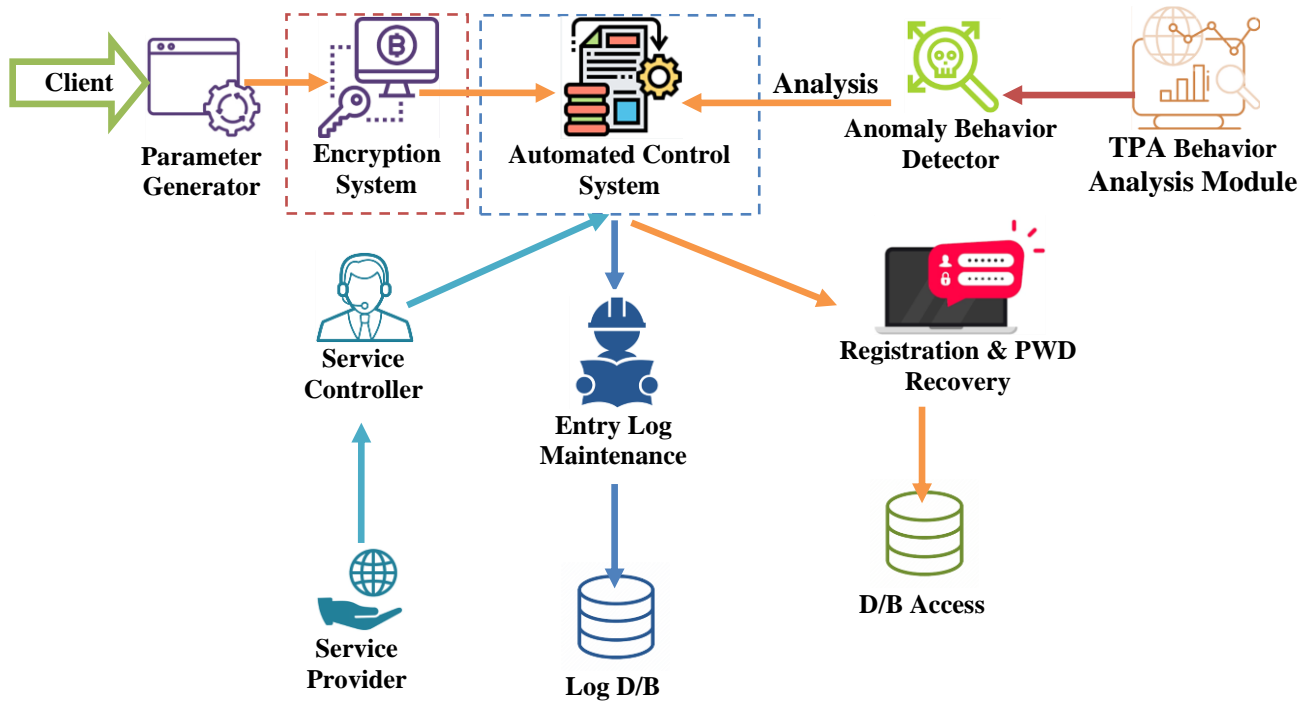

**Fig. 1 Architecture of Proposed Model**

### 4.1. Overview
#### 4.1.1. Register
Motherboard No, Disk No, User Password, 3 Security questions.

#### 4.1.2. Log in
Client types the password, key from the client is formed from Motherboard No, Disk No, User Password, and time. Encrypted and sent to Controller.

#### 4.1.3. Encryption from the Client
Key from the client (RSA derived algorithm - 128-bit cipher)

#### 4.1.4. Automated Controller
Receives the key from the client and intimates CSP to provide the key.

#### 4.1.5. CSP
On request from the automated Controller, sends the key to the Controller.
(Key encrypted by RSA-derived algorithm – 128-bit cipher)

#### 4.1.6. Verification
Automated Controller verifies if the client key and CSP key match.

#### 4.1.7. Log Entry
All attempts are logged by the automated Controller.

### 4.2. Registration Phase
Acquire the disc and motherboard numbers using a client end module all through the registration course. In order to create a processed Motherboard number, we process the alphanumeric motherboard number. The ASCI representation of each character in the practiced Mother Board integer number makes up the number of the Mother Board. In addition, the customer is required to input a few safety measures questions in case a future password recovery is necessary. The database has these specifics. Obtain the user's alphanumeric password as well. A secure key is given to the CSP for message encryption. It should be required to replace this key every three months. All authentications pertaining to the authentication scheme are encrypted using this secure key.

*Algorithm*
A client-end component is launched. The Mother Board identifier could be detected by this unit, which can then processing it to produce the necessary ASCI format. The disc id is found. It gets the passcode from the user. A "Case" would have been calculated.

$$Case = (Disk\_No + Moboard\_Num) * User\_Pwd$$

*Notation*
Disk_No = Disk Id is recognized by the client
Moboard_Num = On client side -> Board number.
User_Pwd = the password which is being selected by the user.
The preparation of this case took 30 microseconds.

### 4.3. Dynamic Nature of Encrypted UEG
To improve security, the Encrypted UEG is time-bound as well as reactive. It takes 120 seconds to finish the entire transaction. The current OTP method has a five-minute minimal waiting period. A safety feature named "Time Case" is created. Time Case is formed by using Cases, Minutes, and Hours as the parameters.

1) Processed Min = (Minutes + Primes) * (subsequent primes) ^2 (prime number < 10, to increase the processing capability in sluggish networks)
2) Processed Hour = Processed Hour + (Processed Min * 2^n) (where the condition 1<=n <=6)
3) Time Case = Processed Min * Processed Hour * Case

While analyzing the time case on great velocity ISDN connections is possible; we often choose lower repetition bounds for demonstrations on sluggish networks. Now, the Time Case created is encoded with Symmetric encryption. The Time Case is encrypted using the selected secret key.

#### 4.3.1. Merits
RSA is currently among the most security awareness in use. Furthermore, we extend the method's inconsistency by a further step. The TimeCap is ever-evolving. The parameter never repeats and changes over time. It is also RSA encrypted. Because TimeCap depends on Minute, Hour, and Capsule, it is protected from spying even if the RSA key is leaked ( DiskNo, MotherBoard No, UserPassword )

#### 4.3.2. Verification
On the side of the (Automated Controller) Third party, the encrypted TimeCap (initiator) is received. The TimeCap would be passed from the third party to the cloud service provider [42-44]. Next, the cloud service provider would examine the administrator-set right of entry for the client. The cloud service provider on its end would then produce the "key." This secret key is sent to the automatic supervisor to allow the client to utilise the public cloud. A log entry is created every time a user tries to log into the system.

### 4.4. TPA Behavior Analysis
Every audit an auditor completes is continuously monitored. The intricacy of the data segments is categorized. The widespread belief is that audit duration increases in direct proportion to audit complexity. An auditor is placed on a blocklist if it is discovered that their audit durations differ from the overall patterns of other auditors' audit durations over time [45]. To confirm accuracy, the data segments are audited once more. The good news is that the system considers the dynamic changes in audit patterns, and those auditors cannot predict how long an audit will take for a particular data segment. This secret key is transmitted to the automatic supervision so the customer can use the cloud service.

As the designer perfects their depiction of programming, they finish the comprehensive schematic, which tackles many areas of the prototype design. There is a bridge to the requirements in the information flow diagram. This cross-goal reference is to prove that the design phase satisfies each requirement. Specifying certain elements are essential for carrying out certain criteria.

## 5. Performance Measure and Optimization

According to research on cyber attacks on Sony Corporation and iCloud, upgrading the cloud gateway with a highly complex approach has become necessary to increase the data's security. The institution's server had the EUEG experimentally installed, and five client nodes with Windows 7 operating system, 8 GB of RAM, and an i5 processor were chosen arbitrarily. Upon signing in, the client node started the EUEG client module. Based on the motherboard number, disc integer, user-assigned password, login time (Minute), and hour of login, a EUEG Key was generated. On the client module, the parameter was encrypted using an RSA-derived technique. Later, for authentication, this was transmitted to the Master server. After obtaining the EUEG from either the customer end, the Central server instructed the CSP component to synthesize the EUEG (secret key) on its side.

The EUEG that the Client provided and the CSP were compared. In less than two minutes, this was confirmed. As a means of authentication, the EUEG pointed the customer to their Google account after verification. However, the authentication failed when we attempted to enter the system via a false attribute. Another effort by EUEG to get into the system failed after the authentication process was prolonged for more than 2 minutes, suggesting that the time window had run out. Various login behavior was used, and the outcomes were all preserved in a repository.

These outcomes were afterwards utilized to create their own training data for the Random Forest Algorithm to identify customer, cloud service provider, and TPA abnormal behavior using their respective training data if any unusual activities lead to keeping a record of entries that were kept on the Central Server, with the Client, CSP, and TPA only being able to see them. Four volunteers who served as auditors were used to test the system and implement TPA behavioral patterns. They were given files and instructed to use a simulator to check the checksums.

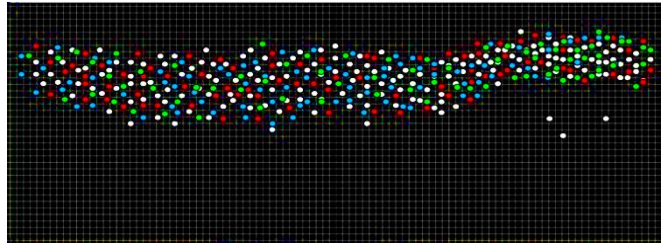

**Fig. 2 Capture the Behavior of various Third Party**

**Table 2. Capture Behavior of Various Third Party**

| Amount of Clients | The typical capture periods (ms) |
|---|---|
| Single Client | 102.18 |
| Multiple amounts of Clients | 91.8 |

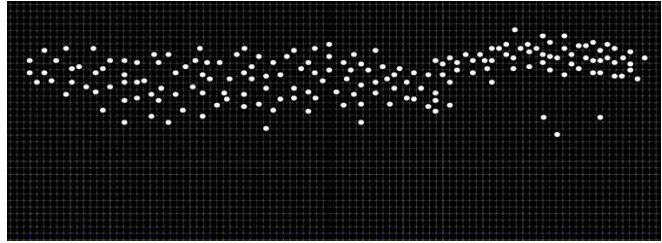

**Fig. 3 Behavior of Blacklist Third Party**

**Table 3. Transfer Times of Client Desktop**

| Amount of Clients | The typical capture periods (ms) |
|---|---|
| Single Client | 44.42 |
| Multiple amounts of Clients | 0 |

In order to display the results in a graphical format, a custom simulator was constructed that included JFreeChart.

The simulator displayed longer checksums for more complicated files and smaller ones for a little less complicated ones. Based on the complexity of the file and the amount of time required to check it, a graph was formed, and the quadrant assessors were assigned the colours Green, Red, White, and Blue. It was discovered that for complex files, the duration significantly increased. The white assessors did deviate from prevailing norms toward the end, nevertheless. When we re-audited a few files chosen by the white assessors, we discovered that three out of them had been audited quickly. Due to this, the audit's intent was harmed, and he was blacklisted, as displayed in Figures 2 & 3. Tables 2 and 3 lists the captured behavior of various third parties and the transfer times of client desktops.

### 5.1. Computation Cost
The customer side, the third-party auditor side, and the cloud service provider side are included in the calculation of the anticipated system's computing cost. To ensure that the proposed system's implementation is thoroughly evaluated to accomplish protection and efficiency using the suggested approaches.

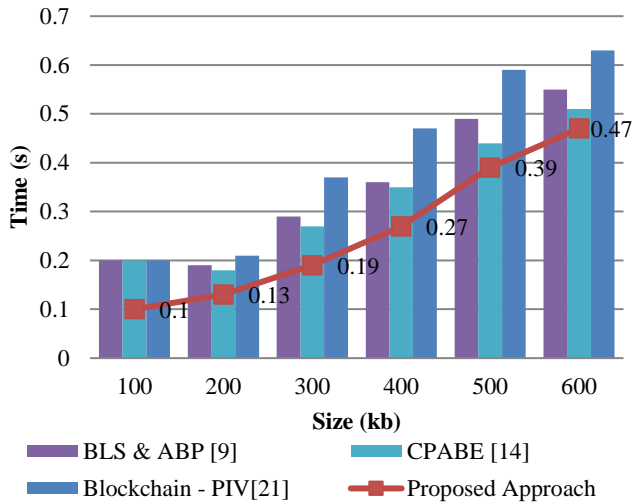

**Fig. 4 Computational Cost of Encryption Process**

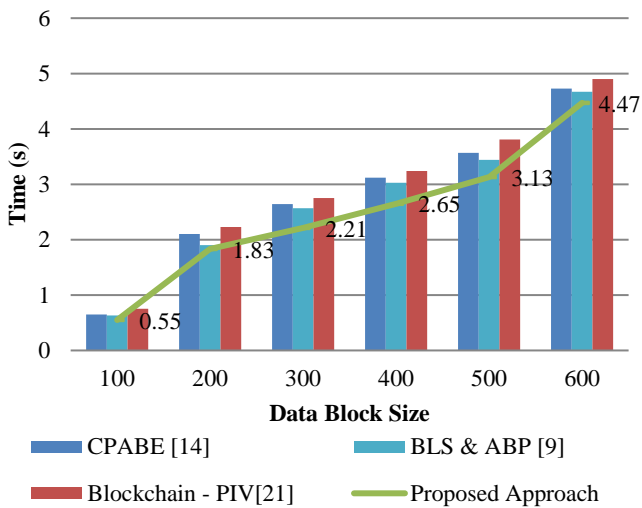

**Fig. 5 Upload Comparison of the block**

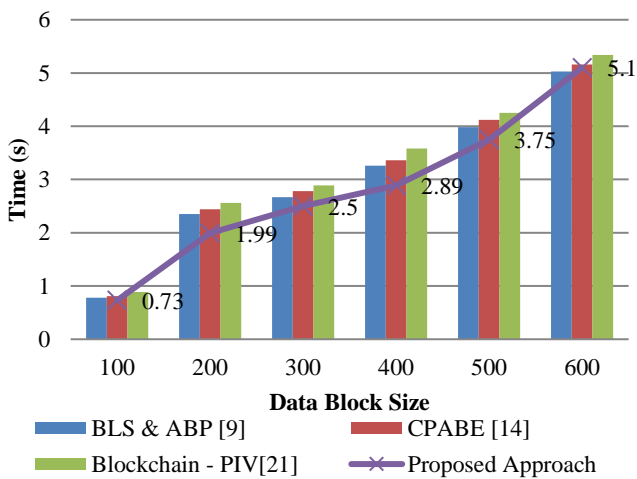

**Fig. 6 Download Comparison of the block**

According to BLS & ABP [9], CPABE [14], and Blockchain – PIV [21], Fig. 4 compares the computational cost of user end encoding for various quantities of anticipated system data. In contrast to BLS & ABP [9], CPABE [14], and Blockchain - PIV [21], the proposed method encrypts 500 KB of data in less than 0.40 seconds, as shown in Fig. 4. Because it provides superior data security for outsourcing and does not require user resources, the computational cost of the projected system's cryptographic expenses is yet appealing. This ensures the proposed system's effectiveness.

### 5.2. Communication Cost

The Communication costs arise from transferring the data to the cloud, retrieving data from the cloud, and responding to auditing challenges. The suggested approach contains randomized chunks that the TPA chooses to argue so because an independent auditor only needs multiple uploads, and the expense is dependent on the chunks that the TPA has already sent for the inspection and the subsequent for its reaction. We run some experiments to determine how long it takes to upload and retrieve files from the cloud. After breaking the data into a number of blocks, we encode the resultant frames before sending them to the cloud server. Figures 5 and 6 show the outcomes of different approaches for downloading and uploading as a result.

We can see from Figs. 5 and 6 that with small block sizes, the average communication time is steady. However, the load builds up gradually. We also get the conclusion that it takes less time to upload data blocks than it does to download them. As a result, the time needed to upload and download blocks from the cloud grows exponentially as the amount of blocks increases. The block must be downloaded from the cloud at a communication overhead of (0.61) seconds. With much better performance measurement, our proposal provides security for the data stored in the cloud. The overall system analysis and design deliver the simple key generation with many intricate ideas to prevent intruders from getting into the cloud gateway. This project is greatly improved by the Third Party Behaviour (TPA) Analyzer, allowing the user to learn about third parties as well as the data files they submit to the disc.

## 6. Conclusion

Continuous security checks are required for data that has been outsourced to a possibly unreliable party. We refer to the integrity and privacy of remotely stored data as security. The cloud service provider (CSP) may make any changes or make storing provisions for users' data if there is no familiar data authenticity checking method. In order to prevent data from being leaked to an unauthorized party, the authenticity verification process we used had to be done with extreme care. Users can designate a TPA to confirm the validity of their data in a public auditing system. However, this delegation raises privacy issues because the TPA may be able to determine data blocks from various server replies to his request for verification.

So, we have improved cloud security by using static system settings and time to provide a dynamic key that is further encrypted. The technique is reliant on hardware specifications that are unique to each computer. Auditor activity is also watched out for. The log entry module keeps track of how the overall system is operating. The institution's servers have the suggested system set up. We added the encrypted UEG to the Google Developers API list, enabling our Java web application to access Google Drive (Google Inc.'s cloud division). However, we plan to use Google Drive to create the behavior module for the TPA in real-time by utilizing the Random Forest Algorithm that is currently being prepared. According to a thorough performance and security assessment, the proposed system is exceptionally secure and efficient.

## References

[1] Chen, Y., Li, M., Huang, X., and Wang, L, "Blockchain-Based Data Storage: A Survey," *Journal of Parallel and Distributed Computing*, vol. 123, pp. 24-41, 2017.

[2] Junfeng Xie et al., "A Survey of Blockchain Technology Applied to Smart Cities: Research Issues and Challenges," *IEEE Communications Surveys & Tutorials*, vol. 21, no. 3, pp. 2795-2830, 2019. [CrossRef] [Google Scholar] [Publisher link]

[3] S.M. Udhaya Sankar et al., "Safe Routing Approach By Identifying and Subsequently Eliminating the Attacks in MANET," *International Journal of Engineering Trends and Technology*, vol. 70, no. 11, pp. 219-231, 2022. [CrossRef] [Publisher link]

[4] D. Dhinakaran, and P. M. Joe Prathap, "Preserving Data Confidentiality in Association Rule Mining Using Data Share Allocator Algorithm," *Intelligent Automation & Soft Computing*, vol. 33, no. 3, pp. 1877–1892, 2022. [CrossRef] [Google Scholar] [Publisher link]

[5] Udhaya Sankar S.M et al., "Mobile Application Based Speech and Voice Analysis for COVID-19 Detection Using Computational Audit Techniques," *International Journal of Pervasive Computing and Communications*, vol. 18, no. 5, pp. 508-517, 2022. [CrossRef] [Google Scholar] [Publisher link]

[6] Guofen Lin et al., "An Expressive, Lightweight and Secure Construction of Key Policy Attribute-Based Cloud Data Sharing Access Control," *Journal of Physics: Series*, vol. 910, p. 012010, 2017. [CrossRef] [Google Scholar] [Publisher link]

[7] Xiuqing Lu, Zhenkuan Pan, and Hequn Xian, "An Efficient and Secure Data Sharing Scheme for Mobile Devices in Cloud Computing," *Journal of Cloud Computing*, vol. 9, no. 1, pp. 1–13, 2020. [CrossRef] [Google Scholar] [Publisher link]

[8] Wenting Shen et al., "Enabling Identity-Based Integrity Auditing and Data Sharing with Sensitive Information Hiding for Secure Cloud Storage," *IEEE Transactions on Information Forensics and Security*, vol. 14, no. 2, pp. 331–346, 2019. [CrossRef] [Google Scholar] [Publisher link]

[9] Baidaa Abdulrahman Jalil et al., "A Secure and Efficient Public Auditing System of Cloud Storage Based on BLS Signature and Automatic Blocker Protocol," *Journal of King Saud University – Computer and Information Sciences*, vol. 34, no. 7, pp. 4008–4021, 2022. [CrossRef] [Google Scholar] [Publisher link]

[10] Dnyanada N. Meshram, Shrikant Zade, and Leena Patil, "Privacy Preserving Implementing for TPA Data Sharing in Cloud," *International Journal of Engineering Science and Computing,* vol. 11, no. 5, pp. 28025 - 28029, 2021.[Google Scholar] [Publisher link]

[11] Xinrui Ge et al., "Towards Achieving Keyword Search Over Dynamic Encrypted Cloud Data with Symmetric-Key Based Verification," *IEEE Transactions on Dependable and Secure Computing*, vol. 18, no. 1, 2021. [CrossRef] [Google Scholar] [Publisher link]

[12] Yi Sun et al., "An Adaptive Authenticated Data Structure with Privacy-Preserving for Big Data Stream in Cloud," *IEEE Transactions on Information Forensics and Security*, vol. 15, pp. 3295–3310, 2020. [CrossRef] [Google Scholar] [Publisher link]

[13] Nouha Oualha et al., "Lightweight Attribute-Based Encryption for the Internet of Things," *2016 25th International Conference on Computer Communication and Networks*. [CrossRef] [Google Scholar] [Publisher link]

[14] Xiaolongxu et al., "Multi-Authority Proxy Re-Encryption Based on CPABE for Cloud Storage Systems," *Journal of Systems Engineering and Electronics*, vol. 27, no. 1, pp. 211 – 223, 2016. [Google Scholar] [Publisher link]

[15] Cong Wang et al., "Privacy-Preserving Public Auditing for Data Storage Security in Cloud Computing," *Proceedings IEEE INFOCOM*, pp. 1–9, 2010. [CrossRef] [Google Scholar] [Publisher link]

[16] Yuan Ping, "Public Data Integrity Verification Scheme for Secure Cloud Storage," *Information*, vol. 11, no. 9, p. 409, 2020. [CrossRef] [Google Scholar] [Publisher link]

[17] Guofeng Lin et al., "A Collaborative Key Management Protocol in Ciphertext Policy Attribute-Based Encryption for Cloud Data Sharing," *IEEE Access*, vol. 5, pp. 9464 – 9475, 2017. [CrossRef] [Google Scholar] [Publisher link]

[18] D. Dhinakaran et al., "Recommendation System for Research Studies Based on GCR," *International Mobile and Embedded Technology Conference (MECON)*, pp. 61-65, 2022. [CrossRef] [Google Scholar] [Publisher link]

[19] Jena Catherine Bel D et al., "Trustworthy Cloud Storage Data Protection Based on Blockchain Technology," *2022 International Conference on Edge Computing and Applications (ICECAA)*, pp. 538-543, 2022. [CrossRef] [Google Scholar] [Publisher link]

[20] K. Sudharson et al., "Hybrid Deep Learning Neural System for Brain Tumor Detection," *2022 2nd International Conference on Intelligent Technologies* (CONIT), pp. 1-6, 2022. [CrossRef] [Google Scholar] [Publisher link]

[21] Yuan Zhang et al., "Blockchain-Based Public Integrity Verification for Cloud Storage against Procrastinating Auditors," *IEEE Transactions on Cloud Computing*, vol. 9, no. 3, pp. 923–937, 2021. [CrossRef] [Google Scholar] [Publisher link]

[22] Wu, C. H., Lin, I. C., and Chou, C. F, "Cloud Storage Auditing With Blockchain and Merkle Tree," *Journal of Ambient Intelligence and Humanized Computing*, vol. 11, no. 2, pp. 817-826, 2020.

[23] T. Sujithra et al., "Id Based Adaptive-Key Signcryption for Data Security in Cloud Environment," *International Journal of Advanced Research in Engineering and Technology* (IJARET), vol. 11, no. 4, pp. 167-182, 2020. [Google Scholar] [Publisher link]

[24] S. M. Udhaya Sankar et al., "Efficient Data Transmission Technique for Transmitting the Diagnosed Signals and Images in WBSN," *4th International Conference on Recent Trends in Computer Science and Technology*, pp. 251–256, 2022. [CrossRef] [Google Scholar] [Publisher link]

[25] Haiping Huang et al., "A Blockchain-Based Privacy-Preserving Cloud Storage System with Consensus Mechanisms," *Computer Security*, vol. 119, pp. 433-446, 2021. [CrossRef] [Publisher link]

[26] Lee, C. H., Kim, J. M., and Lee, K. J, "A Blockchain-Based Multi-Party Computation Approach for Compliance in Cloud Storage," *Journal of Parallel and Distributed Computing*, vol. 157, pp. 18-29, 2022.

[27] G. Gomathy et al., "Automatic Waste Management Based on Iot Using a Wireless Sensor Network," *2022 International Conference on Edge Computing and Applications*, pp. 629-634, 2022. [CrossRef] [Google Scholar] [Publisher link]

[28] S. M. Udhaya Sankar, Mary Subaja Christo, and P. S. Uma Priyadarsini, "Secure and Energy Concise Route Revamp Technique in Wireless Sensor Networks," *Intelligent Automation and Soft Computing*, vol. 35, no. 2, pp. 2337–2351, 2023. [CrossRef] [Google Scholar] [Publisher link]

[29] D. Dhinakaran et al., "Secure Android Location Tracking Application with Privacy Enhanced Technique," *2022 Fifth International Conference on Computational Intelligence and Communication Technologies,* pp. 223-229, 2022. [CrossRef] [Google Scholar] [Publisher link]

[30] Xiong, H., et al., "A Blockchain-Based Cloud Storage System with Proof of Retrieve," *2017 IEEE International Conference on Communications*, pp. 1-6, 2017.

[31] M E Purushoththaman, and Bhavani Buthtkuri, "Effective Multiple Verification Process Ensuring Security and Data Accuracy in Cloud Environment Storage," *SSRG International Journal of Computer Science and Engineering*, vol. 6, no. 7, pp. 1-4, 2019. [CrossRef] [Publisher link]

[32] Dhinakaran D, and Joe Prathap P. M, "Protection of Data Privacy From Vulnerability Using Two-Fish Technique With Apriori Algorithm in Data Mining," *The Journal of Supercomputing*, vol. 78, no. 16, pp. 17559–17593, 2022. [CrossRef] [Google Scholar] [Publisher link]

[33] P. Kirubanantham et al., "An Intelligent Web Service Group-Based Recommendation System for Long-Term Composition," *The Journal of Supercomputing*, vol. 78, pp. 1944–1960, 2022. [CrossRef] [Google Scholar] [Publisher link]

[34] T. Sujithra et al., "Survey on Data Security in Cloud Environment," *International Journal of Advanced Research in Engineering and Technology*, vol. 11, no. 4, pp. 155- 166, 2020. [Google Scholar] [Publisher link]

[35] Li, J et al., "Decentralized and Secure Access Control of Cloud Storage Based on Blockchain and Zero-Knowledge Proof," *IEEE Transactions on Cloud Computing*, 2020.

[36] Laxman Dande, and Soppari Kavitha, "Novel Framework for Public Auditing with Privacy Preserving in Cloud," *SSRG International Journal of Computer Science and Engineering*, vol. 2, no. 7, pp. 31-34, 2015. [CrossRef] [Google Scholar] [Publisher link]

[37] Sankar, S.M., Vijaya Chamundeeswari, , and Jeevaa Katiravan, "Identity Based Attack Detection and Manifold Adversaries Localization in Wireless Networks," *Journal of Theoretical and Applied Information Technology*, vol. 67, pp. 513-518, 2014. [Google Scholar] [Publisher link]

[38] Zhang, Z et al., "A Blockchain-Based Approach for Secure and Efficient Data Sharing in Cloud Storage," *IEEE Transactions on Services Computing*, vol. 14, no. 3, pp. 544-558, 2021.

[39] D. Dhinakaran et al., "Mining Privacy-Preserving Association Rules Based on Parallel Processing in Cloud Computing," *International Journal of Engineering Trends and Technology*, vol. 70, no. 30, pp. 284-294, 2022. [CrossRef] [Google Scholar] [Publisher link]

[40] Sankar, S.M.U., Revathi, S.T., and Thiagarajan, R., "Hybrid Authentication Using Node Trustworthy to Detect Vulnerable Nodes," *Computer Systems Science and Engineering*, vol. 45, no. 1, pp. 625–640, 2023.

[41] B. Murugeshwari et al., "Data Mining With Privacy Protection Using Precise Elliptical Curve Cryptography," *Intelligent Automation and Soft Computing*, vol. 35, no. 1, pp. 839–851. [CrossRef] [Google Scholar] [Publisher link]

[42] L. Srinivasan et al., "Iot-Based Solution for Paraplegic Sufferer to Send Signals to Physician Via Internet," *SSRG International Journal of Electrical and Electronics Engineering*, vol. 10, no. 1, pp. 41-52, 2023. [CrossRef] [Publisher link]

[43] D. Selvaraj et al., "Outsourced Analysis of Encrypted Graphs in the Cloud with Privacy Protection," *SSRG International Journal of Electrical and Electronics Engineering*, vol. 10, no. 1, pp. 53-62, 2023. [CrossRef] [Publisher link]

[44] D. Dhinakaran, and P.M. Joe Prathap, "Ensuring Privacy of Data and Mined Results of Data Possessor in Collaborative ARM," *Pervasive Computing and Social Networking*," vol. 317, pp. 431 – 444, 2022. [CrossRef] [Google Scholar] [Publisher link]

[45] Wang, L et al., "A Blockchain-Based Trusted Computing Framework for Cloud Storage Data Protection," *Future Generation Computer Systems*, vol. 125, pp. 221-23, 2022.

[46] Cao, Y., Hu, J., and Dai, Y, "A Privacy-Preserving Blockchain-Based Cloud Storage System with Homomorphic Encryption," *IEEE Access*, vol. 7, pp. 63129-63139, 2019.

[47] Dhinakaran et al., "Assistive System for the Blind with Voice Output Based on Optical Character Recognition," *International Conference on Innovative Computing and Communications,* vol. 492, 2023. [CrossRef] [Google Scholar] [Publisher link]